# The Pattern of Correlated X-ray Timing and Spectral Behavior in GRS 1915+105

Xingming Chen[1,2], Jean H. Swank[3], and Ronald E. Taam[1]

[1]Department of Physics & Astronomy, Northwestern University, Evanston, IL 60208
chen@apollo.astro.nwu.edu, taam@ossenu.astro.nwu.edu

[2]UCO/Lick Observatory, University of California, Santa Cruz, CA 95064

[3]NASA, Goddard Space Flight Center, Greenbelt, MD 20771
swank@pcasun1.gsfc.nasa.gov

## ABSTRACT

From data obtained from the PCA in the 2-11 keV and 11-30.5 keV energy range, GRS 1915+105 is seen during RXTE observations between 1996 May and October on two separate branches in a hardness intensity diagram. On the hard branch, GRS 1915+105 exhibits narrow quasi-periodic oscillations ranging from 0.5 to 6 Hz with $\frac{\Delta\nu}{\nu} \sim 0.2$. The QPOs are observed over intensities ranging from about 6,000 to 20,000 counts s$^{-1}$ in the 2 - 12.5 keV energy band, indicating a strong dependence on source intensity. Strong harmonics are seen, especially, at lower frequencies. As the QPO frequency increases, the harmonic feature weakens and disappears. On the soft branch, narrow QPOs are absent and the low frequency component of the power density spectrum is approximated by a power-law, with index $\sim -1.25$ for low count rates and $\sim -1.5$ for high count rates ($\gtrsim 18000$ cts/s). Occasionally, a broad peaked feature in the 1-6 Hz frequency range is also observed on this branch. The source was probably in the very high state similar to those of other black hole candidates. Thermal-viscous instabilities in accretion disk models do not predict the correlation of the narrow QPO frequency and luminosity unless the fraction of luminosity from the disk decreases with the total luminosity.

*Subject headings:* accretion, accretion disks — binaries: close — black hole physics — stars: individual (GRS 1915+105, GS 1124-68, GX 339-4) — X-rays: stars

## 1. INTRODUCTION

The X-ray transient source, GRS 1915+105, was discovered in 1992 August by the WATCH all sky monitor instrument on GRANAT (Castro-Tirado, Brandt, & Lund 1992). The combination of the observations using the BATSE experiment on CGRO and the SIGMA telescope on GRANAT reveal that GRS 1915+105 is highly variable in the 20-230 keV range as reported by Harmon et al. (1992, 1994) and Greiner et al. (1994) and in the 35-250 keV range by Finoguenov et al. (1994). More recently, erratic variability has been observed by Greiner, Morgan, & Remillard (1996), who have shown from RXTE observations in 1996 that the X-ray intensity of GRS 1915+105 has extreme variations on time scales ranging from sub-seconds to days. Power law fits to the source spectrum during the outburst in 1992 indicated significant softening over a four month time interval with the spectral index evolving from -2 to -2.8. The detection and observation of GRS 1915+105 at lower energies in the 1-10 keV band by ASCA (Nagase et al. 1994; Ebisawa, White, & Kotani 1995), and in the 0.1-2.4 keV band by ROSAT (Greiner et al. 1993) has provided an estimate of the hydrogen absorption column density to the source of $\sim 5 \times 10^{22} \text{cm}^2$ (Greiner et al. 1994; Ebisawa et al. 1995).

GRS 1915+105 is also unusual in showing superluminal plasmoid ejections (Mirabel & Rodriguez 1994). It was the first transient X-ray source to exhibit jetlike emission and relativistic ejection. Since its discovery, only one other hard X-ray transient source in the Galaxy, GRO J1655-40, has been confirmed to exhibit relativistic ejection of plasmoids (Harmon et al. 1995; Hjellming & Rupen 1995). A sequence of plasmoid ejections were detected by the VLA during 1994 March and April (Mirabel & Rodriguez 1994), indicating apparent superluminal motion of two radio jets. Based on hydrogen absorption line measurements at 21 cm, a distance of $d \approx 12.5 \pm 1.5$ kpc has been inferred, which is consistent with the high reddening ($A_V = 28$ magnitudes) inferred from the X-ray absorption studies.

GRS 1915+105 has also been studied in the optical wavelength region. A probable counterpart has been detected by Böer, Greiner, & Motch (1996) only in the I band, at 23.4 mag. The infrared spectroscopy during the 1994 outburst showed conspicuous H and He emission lines (Castro-Tirado, Geballe, & Lund 1996). Although GRS 1915+105 has been discussed as a black hole candidate source (BHC), there is no general consensus on the nature of GRS 1915+105 and its possible companion.

RXTE (Rossi X-ray Timing Explorer) started monitoring GRS1915+105 with the All Sky Monitor (ASM) on 1996 February 21. The source was discovered to have renewed its activity (Morgan & Remillard 1996). Interesting results from the ASM and Proportional Counter Array (PCA) have been reported by Greiner et al. (1996), Morgan & Remillard (1996), and Thomas et al. (1996).

In this *Letter* we present new results from spectral and temporal analysis of PCA data taken between 1996 May and October. We focus on the features of QPOs with centroid frequencies ranging from 0.5 - 6 Hz and on the discovery of two separate branches in the hardness intensity domain in GRS 1915+105. A QPO at 2.64 Hz was discovered in this range by Thomas et al. (1996). In some observations (Morgan & Remillard 1996; Morgan, Remillard, & Greiner 1997) oscillations at lower frequencies are prominent which have a different character and in some observations a higher frequency QPO ($\sim 67$Hz) is seen. We do not address these here.

## 2. OBSERVATIONS AND RESULTS

The PCA consists of five nearly identical xenon-filled proportional counter units (PCUs) each with collecting area of 1250 cm$^2$ (see Jahoda et al. 1996). The detectors are sensitive to X-rays of energy band 2-60 keV with 17% energy resolution at 6 keV. For comparison, the Crab produces a count rate (summed over the 5 detectors) of 12500 cts/s, and background of $\sim 110$ cts/s.

Starting in April the PCA and HEXTE observed GRS1915+105 twice a week. Discovery of its activity triggered target of opportunity observations for several accepted observing programs with limited observing times. Data for other observations is in the RXTE public archive. When the source appeared to be declining a series of observations was made for our proposal for study of QPO during the decay of transient outbursts of BHCs. These were interleaved with the public observations from July 23 through September 25. For the analysis, we used the FTOOLS supported by the RXTE Guest Observer Facility and the HEASARC Xronos and Xspec. Each observation covered 2-3 satellite orbits with each orbit consisting of $\sim 2500$s of data and a no-data interval of $\sim 1500$s due to Earth occultation and/or South-Atlantic Anomaly passage. Throughout the observation, the PCA count rate (cts/s) in the energy band of 2-11 keV was above 5000, and in the energy band of 11-30.5 keV was above 400. The corresponding background count rates are less than a few percent of the source count rate.

The hardness-intensity diagram of GRS 1915+105 is shown in Figure 1 for selected dates of observation. The hardness ratio is calculated by using the PCA Standard2 mode which has a bin size of 16 s and a total energy band of 2-60 keV. While a very wide range of behavior is indicated, there are at least two distinct branches in which there are characteristic properties. These branches are distinguished by the amplitude and time scales of variability seen in the light curve. In particular, the hard branch (July 11, 29; August 3, 25, 31; September 7) extends over a hardness of 0.1 to 0.2 and corresponds to a variability in intensity by $\sim \pm 10\%$ about the mean for time scales from minutes to an hour, although this type of behavior exists for intensities covering a wide range. On the other hand, a soft branch is suggested on which the hardness is in the range 0.05 to 0.08. At the low count rate end of the soft branch (May 5, 14, 29; June 5; July 3), the source has minute to hour variations of count rates about $\pm 25\%$, which change the hardness by 10%. May 5 and May 14 appear to have hardness/intensity characteristics which extrapolate from those of the other observations in this group. These are the observations in which strong and sharp very low frequency oscillations were reported (Morgan & Remillard 1996). The ASM data in Greiner et al. (1996) suggest, and the PCA data confirm, that the observations between MJD 50179 (April 6) and 50202 (April 29) would lie on the soft branch. However, this part of phase space for the source clearly had very complex temporal behavior which required separate analysis (Morgan et al. 1997). At the high count rate end of the soft branch (May 26 and September 11) the source varies by $\sim 100\%$. These observations represent a distinct state of extreme instability in comparison to the more stable low intensity parts of this branch. During the observation of May 26, for 2-30.5 keV count rates below 18000 counts s$^{-1}$, this branch bifurcates. This is caused by a harder energy spectrum during the dips of the light curve (Swank, Chen, & Taam 1997). On the main section (i.e. count rate higher than 18000 cts/s), the hardness variation is between 0.07 and 0.08.

The properties of the short time-scale variations are distinct on each branch. The QPOs observed on four different dates when the source is on the hard branch are illustrated in the power density spectrum in Figure 2. For the power spectrum, we have used the PCA B_2ms_4B_0_35_H mode which has a bin size of $2^{-9}$ s and an energy band of 2-12.5 keV. We use three bin averages (0.005859375 s) for the FFT results we show in this paper. Power spectra with the normalization used by Leahy (Leahy et al. 1983) were calculated and divided by the mean rate for a density with units of $(rms/mean)^2 Hz^{-1}$. The Poisson noise with a correction for dead time of $10-20\%$ was subtracted. The QPO peaks are relatively narrow ($\frac{\Delta\nu}{\nu} \sim 0.2$). Harmonics are usually present with a strength that decreases with increasing source intensity (see also Fig. 1). The red noise for frequencies lower than the QPO frequency is flat on this branch (note that for September 7, the power decreases slightly between 1 and 4 Hz). The power density spectra of the source on the soft branch can be approximately described by a power law with an index of $\sim -1.25$ at the low count rate end (Fig. 3) and $\sim -1.5$ at the high count rate end (Fig. 4). Figure 3 also indicates that at the low count rate end there is a peaked noise component near 2 Hz, which does not depend on the source intensity. For the data at the highest source intensities (which has an average of more than 28000 cts/s) on 11 September, there is a broad feature at $\sim 5$ Hz and probably a weak harmonic component (see Fig. 4).

The frequency of the QPOs on the hard branch is a sensitive function of the source intensity. Figure 5 shows that the QPO frequency ranges from 0.7 Hz to 6 Hz for intensities ranging from 7000 to only 15000 cts/s in the 2-12.5 keV energy band. Here, if the source intensity changes gradually in an individual observation (such as on August 31 and on September 7), the observation was divided into several shorter time intervals for separate FFT analyses. Through this range of intensities, the width of the QPO is relatively narrow. Above 15000 cts/s, the QPO frequency remains relatively insensitive to count rate at $\sim 5.5 - 6.5$ Hz. For very high and low count rates ($\gtrsim 30000$ cts/s and $\lesssim 7000$ cts/s) there is little evidence for a feature in the power density spectrum.

## 3. DISCUSSION

GRS 1915+105 has been found to exist on two separate branches in a hardness intensity diagram from RXTE observations during the period from 1996 May to October. Taking account of the differences in the temporal behavior, the soft branch includes distinct states. There are strong similarities to behavior observed in other X-ray transients, particularly, BHCs GS 1124-68 (Nova Muscae 1991) and GX 339-4 (Dotani et al. 1996; Miyamoto et al. 1991). On the hard branch the source softens as the intensity increases. This is similar to that observed in GX339-4 in a very high state (Miyamoto et al. 1991) and in GS 1124-68 during its initial rise to its peak X-ray brightness (see Miyamoto et al. 1993; Ebisawa et al. 1994). On the other hand, this is in contrast to the hardness variation observed for bright low mass X-ray binary systems containing a neutron star component along the so called horizontal and normal branches (Z-sources, see van der Klis 1995).

The timing behavior of GRS1915+105 is also strikingly similar to GS 1124-68 in its very high state and to some observations of GX339-4 in a very high state. For example, in cases where a sharp QPO is observed (in the range between 0.5 - 10 Hz), the power spectrum is flat for frequencies lower than the QPO frequency and a harmonic is present. As the X-ray intensity increases the frequency of the QPO increases (although more steeply for GRS 1915+105), and the strength of the harmonic decreases (Tanaka, Makino, & Dotani 1991; Dotani 1991). In cases where the QPO is absent, the power spectra are approximately a power law with power index of $\sim -1$ to $\sim -1.5$ (Miyamoto et al. 1993).

Although there are obvious differences between GRS 1915+105 and GS 1124-68 (for example, the former has jets and remains X-ray active for a long time and the latter has an exponential decay), we conclude that these two sources are similar in nature, that is, GRS 1915+105 is a BHC and currently is in its very high state. We also suggest that, based on the fact that the harmonic of the QPOs disappears near 2.5-3 Hz in GRS 1915+105 and near 8 Hz in GS 1124-68 (Tanaka et al. 1991), and assuming that the frequency is inversely proportional to the mass of the central object, the mass of GRS 1915+105 is approximately 3 times that of GS 1124-68. The scaling argument and the $5 - 7.5 M_\odot$ range deduced by Orosz et al. (1996) for the mass of GS1124-68 imply that the compact object mass of GRS1915+105 could be as large as $22 M_\odot$. This would be consistent with the interpretation of the 67 Hz QPO in terms of general relativistic hydrodynamical oscillations in the accretion disk (Morgan et al. 1997).

There are other parallels, which however, are imperfect. The broad peaked features seen in Figure 3 resemble features seen in Cyg X-1 and GRO J0422+32 (Kouveliotou et al. 1993; Belloni et al. 1996; Cui et al. 1996), although these sources manifest them in states of much lower luminosity. The QPOs with a positive relation between QPO frequency and source intensity and a flat low frequency noise component are similar to some seen in the Z-sources when on the horizontal branch (van der Klis 1995), although the spectral changes do not correspond. We note, however, that the hardness/intensity relation is not limited only to BHCs, since the Z-source Cyg X-2 also shows such variations on its flaring branch (see Kuulkers, van der Klis, & Vaughan 1996). This similarity with the flaring branch in Cyg X-2 would be consistent with GRS 1915+105 being near the Eddington limit. As Greiner et al. (1996) pointed out, at 12.2 kpc the unabsorbed X-ray luminosity exceeds $10^{39}$ ergs/s. Indeed, while the unabsorbed luminosity is sensitive to the spectral model, we found that the absorbed luminosity reaches the Eddington limit for an object of at least $5 M_\odot$.

The variety of QPOs observed from GRS 1915+105 offer the potential for providing an underlying picture of the mechanism responsible for the quasi-periodic variability in BHCs. Models involving magnetic, viscous, thermal, and radiation feedback instabilities in the accretion flow have been suggested to produce intermittent accretion for neutron star systems (e.g., Fortner, Lamb, & Miller 1989; Miller & Lamb 1993). These may apply to BHCs as well. Among the various models, the thermal-viscous instabilities have been amenable to detailed analysis and have been studied to some extent by Lightman & Eardley (1974), Taam & Lin (1984), Lasota & Pelat (1991), and more recently by Milsom, Chen, & Taam (1994), Chen & Taam (1994), and Abramowicz, Chen & Taam (1995). For instabilities of this type the viscous diffusion time of the unstable inner region of the disk (say, around 10 Schwarzschild radii), is identified with the timescale of the QPO variability ($\sim 0.1 - 2$ s). In this picture, the mass flow rate in the inner region of the disk where the X-ray emission is produced, is modulated. However, simple models do not produce the positive correlation of QPO frequency with source intensity. That is, if one assumes that the mass accretion rate is proportional to the source intensity, then QPOs of decreasing frequency are expected for higher source intensities since higher mass accretion rates lead to larger unstable regions resulting in longer viscous diffusion time scales. The observed correlation in GRS 1915+105 could be achieved, however, if the fraction of energy dissipated in the disk itself decreased as the local mass flow rate increased, since then the unstable region of the disk may decrease (Abramowicz et al. 1995). This circumstance may arise, for example, if a greater fraction of accretion energy is dissipated in a corona. By allowing for the deposition of energy in this manner, the width of the unstable region could decrease with increasing accretion rates. Hence, shorter recurrent time scales and higher QPO frequencies would be expected at higher mass accretion rates. Future phenomenological modeling will be required to determine the feasibility of such a hypothesis for the variability in GRS 1915+105.

This research was supported by NASA under grant NAGW-2526 and by the RXTE NRA-1 grant 10258 through the University Space Research Association (USRA) visiting Scientist Program. We thank Jochen Greiner for his careful consideration, as referee, of our different treatments of RXTE data on this source, which improved our paper.

– 8 –Morgan, E. H., Remillard, R. A., & Greiner, J. 1996, ApJ, submitted

Nagase, F., Inoue, H., Kotani, T., & Ueda, Y. 1994, IAU Circ. 6392

Orosz, J. A., Bailyn, C. D., McClintock, J. E., & Remillard, R. A. 1996, ApJ, 468, 380

Swank, J, Chen, X., & Taam, R. E. 1997, AAS 189th meeting

Taam, R. E., & Lin, D. N. C. 1984, ApJ, 287, 761

Tanaka, Y., Makino, F., & Dotani, T. 1991, in Workshop on Nova Muscae 1991, ed. S. Brandt (Lyngby: Danish Space Research Institute), 125

Thomas, B., Corbet, R., Swank, J., & Focke, W. 1996, IAU Circ. 6435

van der Klis, M. 1995, in X-Ray Binaries, ed. W. H. G. Lewin, J. van Paradijs & E. P. J. van den Heuvel (Cambridge: Cambridge University Press), 252
This preprint was prepared with the AAS LATEX macros v4.0.

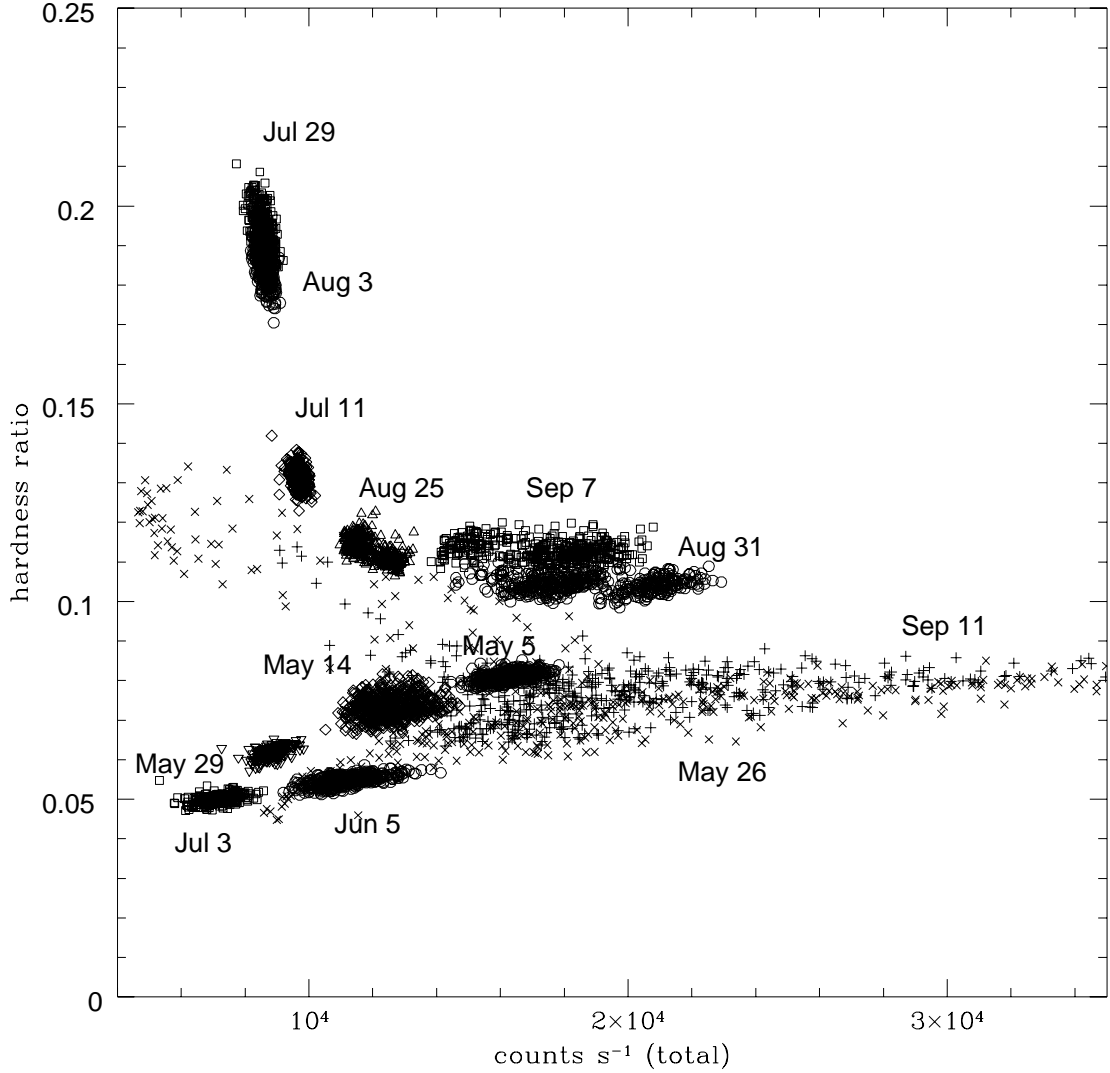

Fig. 1.— The hardness-intensity diagram for the dates indicated. The hardness is defined as the ratio of the count rate in the 11-30.5 keV band to the count rate in the 2-11 keV band. Note that the data appear to define 2 main branches plus an intermediate spur. Some observations are not plotted which would lie nearly on top of others, in particular observations during April which lie near those of May 5 and May 14.

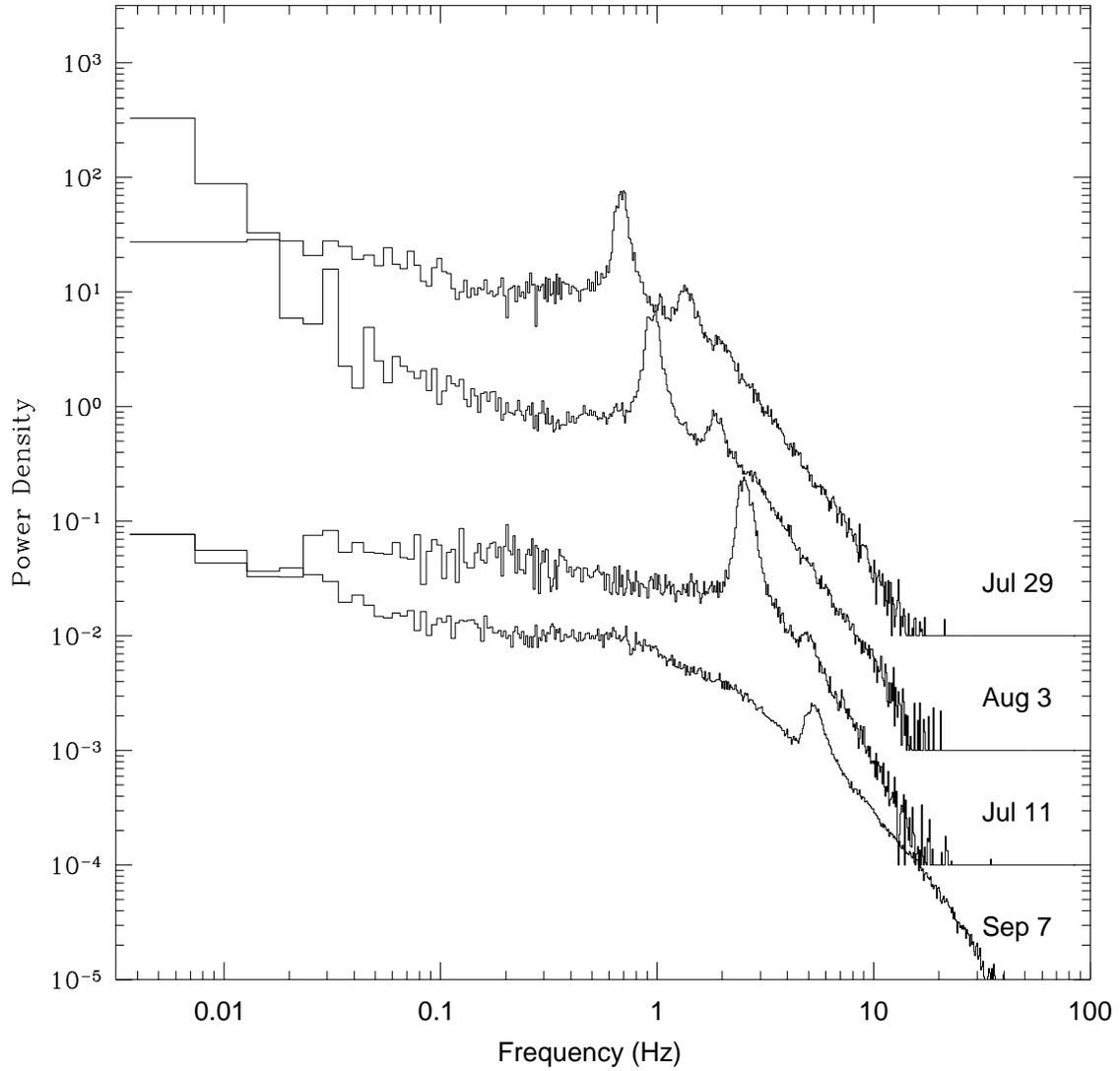

Fig. 2.— Power density spectra for four dates when the source is on the hard branch. Note the sharpness of the features and the existence of harmonics. The spectra are shifted upward by one decade one by one from September 7.

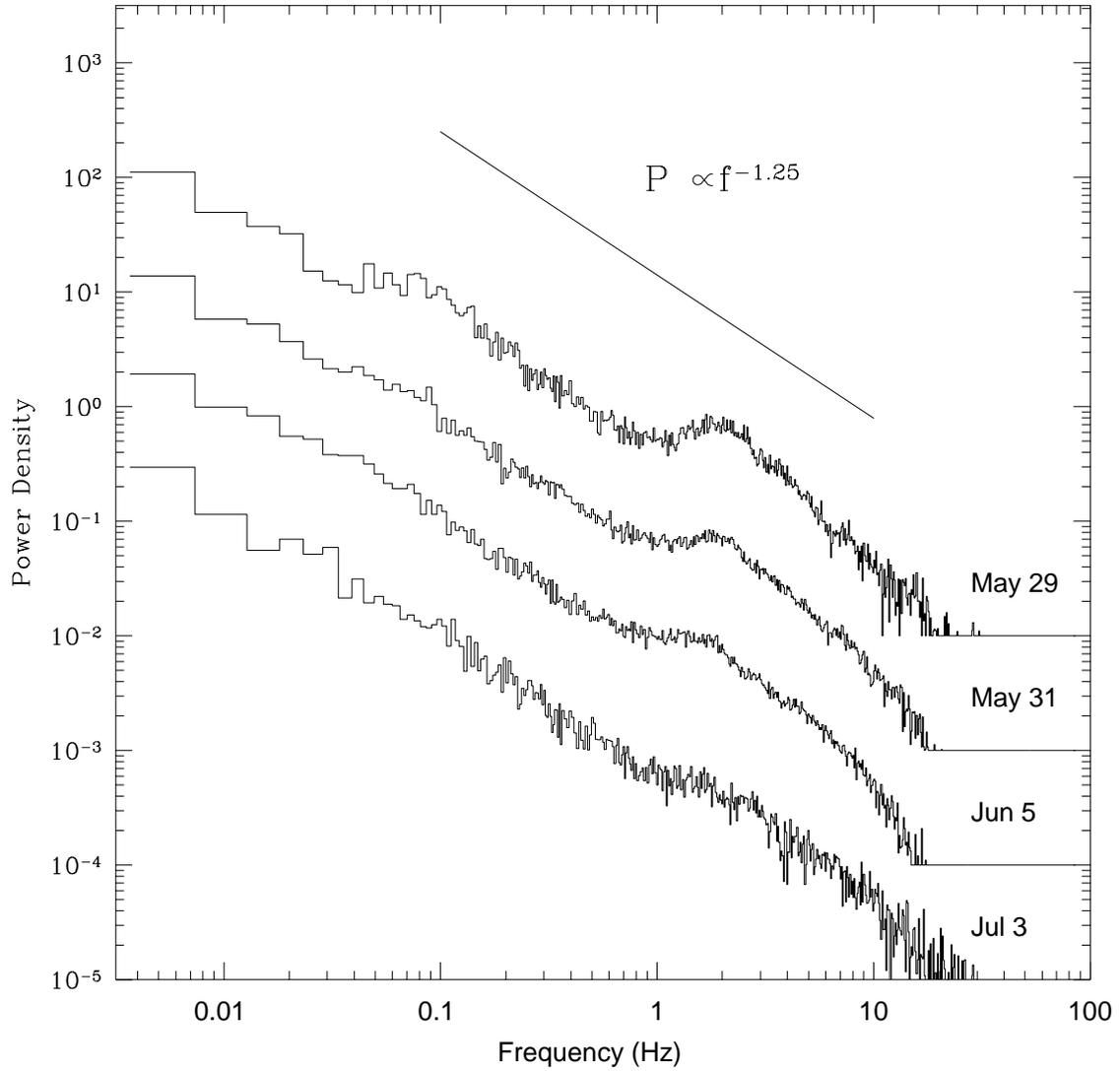

Fig. 3.— Power density spectra for four dates when the source is on the soft branch. Note the broad feature near 2 Hz and the steep shape of the red noise component. The spectra are shifted upward by one decade one by one from July 3.

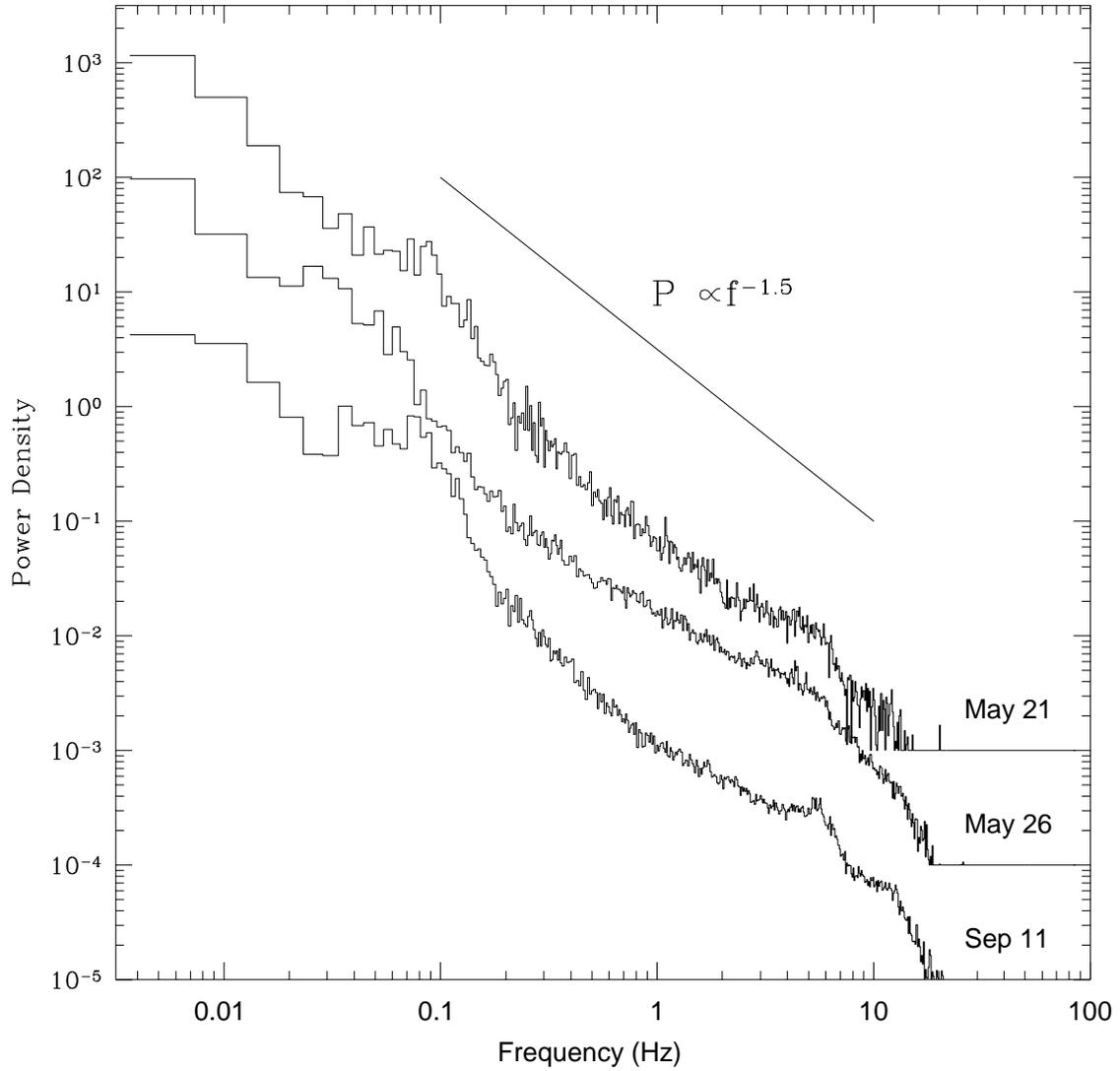

Fig. 4.— Power density spectra for three dates when the source is at high count rates on the soft branch. The features are very broad or nonexistent and the red noise component is steep. The spectra are shifted upward by one decade one by one from September 11.

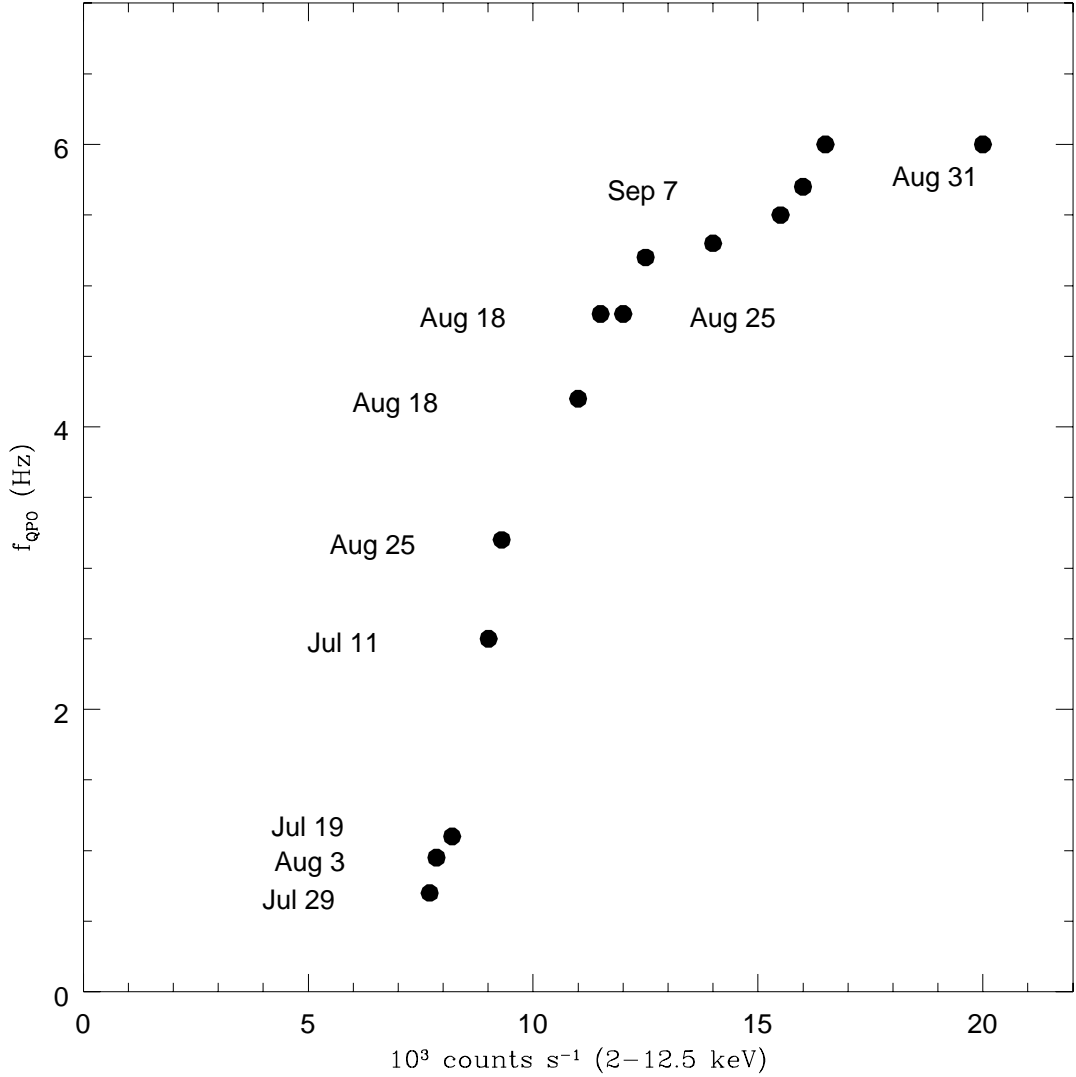

Fig. 5.— The frequency of the features in the power density spectra as a function of source intensity. Fits to the QPO with a Lorenztian give centroid frequencies with errors less than 0.01-0.1 Hz. The count rates for the data used for a power density spectrum varied by less than 500-1000 cts/s. Note the steep dependence between 7000 and 15000 cts/s and the weaker dependence for higher source intensities.